\documentclass[english,11pt,aps,prd,a4paper,preprintnumbers,floatfix,nofootinbib,showpacs,superscriptaddress, notitlepage]{revtex4-1} 

\newcommand{\AddrUNAM}{Instituto de F\'isica, Universidad Nacional Aut\'onoma de M\'exico, A.P. 20-364, Ciudad de M\'exico 01000, M\'exico.}
\usepackage[utf8x]{inputenc} 
\usepackage[]{tikz-feynman}
\usepackage{booktabs}
\usepackage{color}
\usepackage{graphicx}
\usepackage{multirow}
\usepackage{amsmath}
\usepackage{amssymb}
\usepackage[colorlinks=true,citecolor=darkred,urlcolor=darkred, pdfborder={0 0 0}]{hyperref}
\usepackage[normalem]{ulem}
\usepackage{soul}
\definecolor{darkred}{rgb}{0.6,0,0}
\definecolor{drkgrn}{RGB}{0, 51, 0}
\definecolor{gray}{RGB}{128, 128, 128}
\usepackage{soul}

\def\beq{\begin{equation}}
\def\eeq{\end{equation}}

\newcommand{\eps}{\varepsilon}
\def\cevns{CE$\nu$NS~}
\def\u1p{$U(1)^{\prime}$~}
\def\zp{$Z^{\prime}$~}
\def\gp{$g^{\prime}$~}
\def\mzp{$M_{Z}^{\prime}$~}
\setstcolor{red}

\begin{document}

\title{Complementarity between dark matter direct searches and \cevns experiments in $U(1)'$ models}

\author{Leon M. G. de la Vega}\email{leonm@estudiantes.fisica.unam.mx}\affiliation{\AddrUNAM}
\author{L.  J.  Flores}\email{luisjf89@fisica.unam.mx}\affiliation{\AddrUNAM}
\author{Newton Nath}\email{newton@fisica.unam.mx}\affiliation{\AddrUNAM}
\author{Eduardo Peinado} \email{epeinado@fisica.unam.mx}\affiliation{\AddrUNAM}

\begin{abstract}
{\noindent We explore the possibility of having a fermionic dark matter candidate  within $U(1)'$ models  for CE$\nu$NS  experiments in light of the latest COHERENT data and the current and future dark matter direct detection experiments.  A vector-like fermionic dark matter has been introduced which is charged  under $U(1)'$ symmetry, naturally stable after spontaneous symmetry breaking.   We perform a  complementary investigation using CE$\nu$NS experiments and dark matter  direct detection searches to explore dark matter as well as $Z^{\prime}$ boson parameter space.   Depending on numerous other constraints arising from the beam dump,  LHCb,  BABAR,  and the forthcoming reactor experiment proposed by the SBC collaboration,  we explore the allowed region of $Z^{\prime}$  portal dark matter.
 }
\end{abstract}

\maketitle

\section{Introduction}
The coherent-elastic neutrino-nucleus scattering (CE$\nu$NS) has been  observed by the COHERENT collaboration using  cesium-iodide (CsI) detector~\cite{Akimov:2017ade},  almost  40 years after of its first proposal \cite{Freedman:1973yd}.   
The reported results are observed at $6.7^{}\sigma$ significance and are consistent  with the  Standard Model (SM) expectations at $1.5^{}\sigma$~\cite{Akimov:2017ade}. 
Recently,    the first measurement of CE$\nu$NS  using the CENNS-10 liquid argon detector~\cite{Akimov:2020pdx,Akimov:2020czh}  has been reported by  the COHERENT collaboration with greater than $3^{}\sigma$ significance. 
For this process to occur, neutrino energies below $\sim50$ MeV are needed, hence producing nuclear recoil energies of a few keV.  Detecting nuclear recoils with such low energy is the major challenging task for any experiment.

Therefore, the measurement of such low-energy process opens a new window to study and test  the SM at low momentum transfer~\cite{Scholberg:2005qs,Lindner:2016wff,Deniz:2017zok,Miranda:2019wdy},  
 nuclear physics ~\cite{Cadeddu:2017etk,Papoulias:2019lfi,Canas:2019fjw,Cadeddu:2020lky}, as well as neutrino  electromagnetic properties~\cite{Cadeddu:2018dux,Miranda:2019wdy}.   

In recent times,  many studies suggest that  \cevns  is an excellent complement to test new physics beyond the SM,  like sterile neutrinos~\cite{Miranda:2019skf,Blanco:2019vyp,Berryman:2019nvr,Miranda:2020syh}, or even non-standard neutrino interactions~\cite{Liao:2017uzy,Giunti:2019xpr,Esteban:2019lfo,Coloma:2019mbs,Khan:2019cvi, Flores:2020lji}.

However,  here we are interested to investigate  dark matter (DM)  using these  neutrino-nucleus scattering experiments.  It has been known in literature that there exists numerous amount of astrophysical and cosmological evidence for the existence of DM in the universe,   but so far there is no experimental result, neither from direct/indirect detection nor from accelerator-based experiments, that supports its existence. 

Among the many dark matter candidates, the thermally produced Weakly Interacting Massive Particle (WIMP) has received much theoretical and experimental scrutiny.
Many dark matter direct detection experiments, such as the Xenon1T \cite{XENON:2020gfr} or PandaX-II \cite{PandaX-II:2017hlx} experiments for example, have also set stringent limits on the DM - nucleus cross-section.
In particular,   there exists  some interesting  light mediators  models connecting dark photons with the light dark matter, as pointed in 
~\cite{Huh:2007zw, Pospelov:2007mp, Hooper:2008im, Cheung:2009qd, Essig:2010ye, Essig:2013lka,Batell:2014yra}.
Moreover, the COHERENT collaboration has recently studied sub-GeV dark matter models using detector sensitive to \cevns processes in Ref. \cite{COHERENT:2019kwz}.

In this work, we exploit an extension of the SM consisting of an additional anomaly-free $U(1)'$ gauge symmetry with a very light gauge boson, $Z^\prime$, and a dark matter candidate, stable due to a residual symmetry of $U(1)'$.
There exists very well-motivated studies, where a new $U(1)'$ gauge symmetry  arises, namely, in the context of  supersymmetry \cite{Cvetic:1998jxa,Chun:2008by,Frank:2020byg}, grand unified theories \cite{London:1986dk,Hewett:1988xc,Arcadi:2017atc}, and string-motivated models \cite{Cvetic:1995rj,Cleaver:1998gc}.
We will study different scenarios for the $U(1)'$ where dark matter is charged under this symmetry, quarks have   flavor independent charges, while leptons are allowed to have generation-dependent charge, namely $U(1)_{B-L}$, $U(1)_{B-2L_\alpha-L_\beta}$, and $U(1)_{B-3 L_\alpha}$.  In this way the new gauge boson couples to quarks, leptons, and dark matter. 
If the interaction mediates both processes, namely dark matter - nucleus scattering and an extra contribution to neutrino - nucleus scattering, there will be a correlation between both cross-sections. In particular, if an extended gauge symmetry is considered to  mediate both processes, the couplings are also correlated through the gauge coupling and the vector boson mediator mass. 
The light gauge boson will mediate the interaction between dark matter and nuclei and it will also contribute to the interaction of a neutrino with nuclei. We will focus on the complementarity of the searches for light gauge bosons and dark matter from colliders and \cevns experiments.

Furthermore, considering the combined effect of different experimental constraints coming from the COHERENT collaboration,  beam-dump experiments,  the LHCb and the BABAR dark photon searches,  we examine the allowed parameter space to constrain \zp boson and dark matter for each \u1p models. In addition, we also explore the potential of upcoming reactor-based CE$\nu$NS experiment proposed by the   Scintillating Bubble Chamber (SBC) collaboration~\cite{Giampa:2021wte,SBC:2021yal}.
Bounds arising  from astrophysical observations such as Big Bang Nucleosynthesis (BBN) and Cosmic Microwave Background (CMB) have also been shown for the comparison.

We organize this work as follows. Sec. \ref{sec:Cevens} is dedicated to a brief description of \cevns experiments. Dark matter in \u1p models has been discussed in Sec. \ref{sec:DMinU1}. We discuss the DM relic density and direct detection process in Sec. \ref{sec:RDandDD}. Our principal results are illustrated in Sec. \ref{sec:DM-CENES}. Summary of this work and concluding remarks are presented in Sec. \ref{sec:conclusion}.

\section{\cevns processes and \zp boson}\label{sec:Cevens}
The coherent elastic neutrino-nucleus scattering (CE$\nu$NS) was measured by the COHERENT experiment~\cite{Akimov:2017ade} using neutrinos from a stopped-pion source at the Spallation Neutron Source (SNS) at Oak Ridge National Laboratory. The COHERENT collaboration used two different detectors, one made of CsI~\cite{Akimov:2017ade} and the other one made of LAr~\cite{COHERENT:2020iec}. 

Furthermore, there are several experiments trying to measure CE$\nu$NS using reactor antineutrinos, but the need of sub-keV thresholds and great control over background has made this task difficult. One of these experimental proposals is a liquid argon scintillating bubble chamber, currently under construction by the SBC collaboration~\cite{Giampa:2021wte,SBC:2021yal}. With an expected energy threshold of 100~eV, a 10-kg chamber placed near a 1-MW$_{th}$ nuclear reactor will be able to measure SM parameters with high precision, such as the weak mixing angle, in addition to set competitive limits on some beyond SM scenarios~\cite{SBC:2021yal}.

The SM differential cross-section for CE$\nu$NS process is given by~\cite{Drukier:1983gj,Barranco:2005yy,Patton:2012jr}
\begin{equation}
\frac{d\sigma}{dT} = \frac{G_F^2}{2\pi}M_N Q_w^2 \left(2 - \frac{M_N T}{E_\nu^2}\right),
\label{eq:crossSec}
\end{equation}
where $T$ is the nuclear recoil energy, $E_\nu$ is the incoming neutrino energy, and $M_N$ is the nuclear mass. The weak nuclear charge, $Q_w$, is given by \footnote{Neglecting the axial-vector interaction.}
\begin{equation}
Q_w =  Z g_p^V F_Z(Q^2) + N g_n^V F_N(Q^2) \;,
\label{eq:weakCharge}
\end{equation}
where  $g_p^V = 1/2 -2\sin^2\theta_W$ and $g_n^V = -1/2$ are the SM weak couplings, $Q$ is the transferred momentum, $Z(N)$ is the proton (neutron) number of the nucleus,  and $F_{Z(N)}(Q^2)$ the nuclear form factor. Since $g_p^V\sim0.02$, the cross-section depends highly on the number of neutrons $N$.

In presence of a new vector interaction, the cross-section for CE$\nu$NS is affected through the weak nuclear charge (see Eq.~\eqref{eq:weakCharge}) in the following way:
\begin{equation}
Q_{w \,\alpha} =  Z (g_p^V + 2\eps_{\alpha\alpha}^{uV} + \eps_{\alpha\alpha}^{dV})F_Z(Q^2) + N(g_n^V  + \eps_{\alpha\alpha}^{uV} + 2\eps_{\alpha\alpha}^{dV})F_N(Q^2)\;,
\label{eq:weakChargeNSI}
\end{equation}
where $\eps_{\alpha\alpha}^{u(d)V}$ are the parameters that quantify the strength of the new interaction (relative to the weak scale) with $u(d)$ quarks, and $\alpha = (e,\mu,\tau)$. Notice that with this new contribution, the differential cross-section from Eq.~\eqref{eq:crossSec} is now flavor dependent. 
The effective low-energy Lagrangian for the neutrino-quark interactions with a new $Z'$ boson can be written as
\begin{equation}\label{eq:eff_lag}
\mathcal{L}_\mathrm{eff} = -\frac{g'^2}{Q^2 + M_{Z'}^2}\left[ \sum_\alpha x_\alpha \bar{\nu}_\alpha \gamma^\mu P_L \nu_\alpha \right] \left[ \sum_q x_q \bar{q}\gamma_\mu q \right],
\end{equation}
where $g'$ and $M_{Z'}$ are the coupling and mass of the new gauge boson, respectively. Therefore, by comparing this effective Lagrangian with the one from an effective field theory approach, we can relate the $\epsilon^{u(d)V}_{\alpha\alpha}$ parameters with the $Z'$ interaction parameters as
\begin{equation}
\eps_{\alpha\alpha}^{u(d)V} = \frac{g'^2 x_\alpha x_q}{\sqrt{2}G_F (Q^2 + M_{Z'}^2)}\;.
\label{eq:nsiCouplingMass}
\end{equation}

In order to extract limits on the $Z'$ parameters from the measurement by the COHERENT collaboration with the CsI detector, the following $\chi^2$ function can be used~\cite{Akimov:2017ade}
\begin{equation}
\chi^2 = \sum_{i=4}^{15} \left[\frac{N_\mathrm{meas}^i - (1+\alpha)N_\mathrm{th}^i- (1+\beta)B_\mathrm{on}^i}{\sigma_\mathrm{stat}^i} \right]^2 + \left(\frac{\alpha}{\sigma_\alpha}\right)^2 + \left(\frac{\beta}{\sigma_\beta}\right)^2,
\label{eq:chiSquareFunction}
\end{equation}
where $N_\mathrm{meas}^i$ and $N_\mathrm{th}^i$ are the measured and theoretical predicted number of events per energy bin, respectively, $\sigma_\mathrm{stat}^i = \sqrt{N_\mathrm{meas}^i + B_\mathrm{on}^i + 2B_\mathrm{ss}^i}$ is the statistical uncertainty of the measurement, and $B_\mathrm{on}^i (B_\mathrm{ss}^i)$ is the beam-on (steady-state) background. The systematic uncertainties of signal and background normalization are encoded in $\sigma_\alpha = 0.28$ and $\sigma_\beta=0.25$, respectively. The function in Eq.~\eqref{eq:chiSquareFunction} has to be marginalized over $\alpha$ and $\beta$, which are nuisance parameters. Recently it has been shown that the limits for a light vector mediator obtained from the COHERENT measurements with the LAr detector are similar to the ones obtained with the CsI data~\cite{Flores:2020lji,Miranda:2020tif,Cadeddu:2020nbr}. Therefore, in this work we will present bounds only from the CsI measurement.

For the case where there is no current measurement, such as with the SBC-\cevns detector, one can assume the measured signal as the SM expectation plus background. Hence, the projected sensitivities on the  $Z'$ parameters can be obtained with the $\chi^2$ function
\begin{equation}
\chi^2 =\left[\frac{N_\mathrm{meas} - (1+\alpha)N_\mathrm{th}(\gamma)- (1+\beta)B_\mathrm{reac}}{\sigma_\mathrm{stat}}\right]^2 + \left(\frac{\alpha}{\sigma_\alpha}\right)^2 + \left(\frac{\beta}{\sigma_\beta}\right)^2 + \left(\frac{\gamma}{\sigma_\gamma}\right)^2,
\label{eq:chisq_SBC}
\end{equation}
where $B_\mathrm{reac}$ is the background due to the nuclear reactor. Here, the statistical uncertainty is defined as $\sigma_\mathrm{stat} = \sqrt{N_\mathrm{meas} + 4 B_\mathrm{cosm}}$, with $B_\mathrm{cosm}$ the background from muon-induced and cosmogenic neutrons. The nuclear recoil threshold is set to (1+$\gamma$)$\cdot$100~eV, where $\gamma$ is an additional nuisance parameter, while the systematic uncertainties associated to the signal, background, and energy threshold are set to $\sigma_\alpha=0.024$, $\sigma_\beta=0.1$, and $\sigma_\gamma=0.05$, respectively.

\section{Dark matter in a $U(1)'$ extension of the Standard Model}\label{sec:DMinU1}
The $U(1)'$ extension of the SM consists of an extra local gauge symmetry. In our framework the SM fermions are charged under the additional $U(1)'$, while the SM Higgs doublet is left uncharged, as shown in Table \ref{tab:modelcharges}. Scalar singlets  $\phi_i$ are added in order to spontaneously break $U(1)'$ symmetry as given by Table \ref{tab:U1pModels}.
Leptons are charged according to flavor-dependent charges $x_e$, $x_\mu$ and $x_\tau$. Anomaly cancellation conditions restrict the possible set of lepton charges $\{x_e,x_\mu,x_\tau \}$. We consider the solution $\{-1,-1,-1\}$ corresponding to the well-known $B-L$ model, solutions $\{-1,-2,0\}$, $\{-1,0,-2\}$, $\{0,-1,-2\}$ and $\{0,-2,-1\}$ corresponding to $B-2L_\alpha-L_{\beta}$ models with very specific predictions in the leptonic sector~\cite{Flores:2020lji} as well as solutions $\{-2,-1,0\}$, $\{-2,0,-1\}$. Finally,   solutions  $\{-3,0,0\}$, $\{0,-3,0\}$ and $\{0,0,-3\}$ have been considered for the the $B-3L_\alpha$ model~\cite{Heeck:2018nzc,Han:2019zkz,Bauer:2020itv}.
\begin{table}[t!]
    \centering
    $ \begin{array}{|c|c|c|c|c|c|c|c|c|c|c|c|c|c|} \hline
        \text{Symmetry/Field} & Q & u & d & L_e & L_{\mu}& L_{\tau} &e_e & e_{\mu} &e_{\tau} & N_1 &N_2 &N_3 &H \\ \hline
         U(1)'& 1/3 & 1/3 & 1/3 & x_e & x_\mu & x_\tau & x_e & x_\mu & x_\tau& x_e & x_\mu & x_\tau & 0  \\ \hline
    \end{array} $
    \caption{\footnotesize $U(1)^{\prime}$ charges of the model. The charges $x_\alpha$, with $\alpha=e,\mu,\tau$, can take the values $x_\alpha = 0,-1,-2,-3$, while the charges of the quarks are $1/3$.}
    \label{tab:modelcharges}
\end{table}

It is well known that if we extend the SM with a $U(1)'$ symmetry spontaneously broken by a scalar field with a integer charge, it is possible to have a residual symmetry $Z_N$. This happens if there is a scalar or fermion field with fractional $U(1)^\prime$ charge~\cite{Krauss:1988zc}. This mechanism can be responsible for the DM stability and depending on the symmetry it can also tell us if neutrinos are Dirac or Majorana particles~\cite{Bonilla:2018ynb,Bonilla:2019hfb}.
Consider introducing a SM singlet fermion $\chi$ which can be Dirac or Majorana, depending only on its $U(1)^\prime$ charge. To avoid spoiling the anomaly-free nature of the model a vector-like pair of $\chi$, $\chi_L$ and $\chi_R$, is needed. If $\chi_{L/R}$ transform as $1/2$, they will be a pair of Majorana fields and their mass will be provided once the $U(1)^\prime$ is broken by a flavon field $\phi$ transforming as $1$ under $U(1)^\prime$. On the other hand, if $\chi_{L/R}$ transform as $1/3$ they will form a Dirac fermion. Thus, we end up with a residual $Z_3$ symmetry, which stabilizes the DM.

%
For the choice of Dirac fermion charge $1/3$ the resulting dark matter mass eigenstate is given by $\chi=\chi_L+\chi_R$. We can write the dark sector Lagrangian  as follows
\begin{equation}
    \mathcal{L}_{D}= \left( \overline{\chi}\gamma^\mu(\partial_\mu + i \frac{g'}{3} Z_\mu^\prime )\chi  \right) + M_{\chi} \overline{\chi} \chi+h.c. \;
\end{equation}
The only coupling of $\chi$ is therefore to the $Z'$ gauge boson. The relic density of dark matter may be determined by this coupling in the \textit{freeze-out}  regime, through the $Z'$ mediated $\overline{\chi} \chi \rightarrow \overline{f} f $ channel, where $f$ is a SM fermion, as well as the $\overline{\chi} \chi \rightarrow Z^\prime Z^\prime  $ channel. The $Z'$ channel also provides a tree-level spin-independent direct detection signature. It is well known that these two constraints (\textit{freeze-out} relic density and direct detection bounds) are in tension for $m_{\chi}<\mathcal{O}$(10 TeV) in $U(1)'$ models \cite{Escudero:2018fwn,Okada:2018ktp,Han:2020oet,Borah:2020wyc,Kaneta:2016vkq,Alves:2016cqf,Borah:2018smz}. For DM masses below $\mathcal{O}(10)$ GeV  the direct detection bounds are much less stringent, due to poor detector sensitivity to nuclear recoils induced by the light dark matter. For the light dark matter and light $Z'$, the correct relic density is obtained for gauge couplings of order $0.1 - 1$, which are excluded by \cevns experiments. This motivates the use of an annihilation cross-section enhancement mechanism. To avoid enlarging the field content of the model, we consider the resonant $Z'$ mechanism, where $M_\chi \sim M_{Z'}/2$. We parametrize the resonant condition with the $\delta_{\rm  Res}$ parameter defined by the relation
\begin{equation}
M_\chi=\frac{M_{Z'}}{2}\left(1 + \delta_{\rm  Res} \right).
\end{equation}
We implemented the $U(1)'$ models in \texttt{LanHEP} \cite{Semenov:2008jy} and \texttt{micrOMEGAs} \cite{Belanger:2018ccd} to calculate the dark matter observables, scanning over the ranges $(10^{-3} - 50 )$ GeV for the $Z'$ mass, $(10^{-6} - 10^{-1})$ for the $g'$ gauge coupling and  to ensure the resonant annihilation $(0.45-0.55)~M_{Z'}$ for $M_\chi$, varying $|\delta_{\rm  Res}|$ between $(0.001 - 0.1)$.

For the different models we will need different singlet scalar fields $\phi_i$  transforming as $i$ under the $U(1)^\prime$,  as shown in Table \ref{tab:U1pModels}, in order to obtain phenomenologically viable neutrino mass matrices through the type-I seesaw mechanism.
The $Z'$ boson, after $U(1)^\prime$ breaking by the scalars, obtains the mass $M_{Z'}$ for different models as shown in Table~\ref{tab:U1pModels}. The fields we have included are those who give a correct neutrino masses and mixings. For  \textbf{MI} there are no correlations in the active neutrino mass matrix, for  \textbf{MII} there are four cases with good phenomenology and predictions as investigated in ~\cite{Flores:2020lji}, for  \textbf{MIII} there is one correlation, see Appendix \ref{app:MIV} and for \textbf{MIV} there are no correlations.
\begin{table}[t]
    \centering
    \begin{tabular}{|c|c|c|c|}\hline
    & $U(1)^\prime$ models &Scalar Fields& Masses of $Z^\prime$ \\
    & & &$(M_{Z'}^2)$ \\
    \hline \hline
   \textbf{MI} & $U(1)_{B-L}$ & $\phi_2$& $g'^2(4v_2^2)$\\\hline
   \textbf{MII} & $U(1)_{B-2L_{\alpha}-L_{\beta}}$ & $\phi_1$, $\phi_2$&$ g'^2(v_1^2+4v_2^2)$\\\hline
   \textbf{MIII} &  $U(1)_{B-2L_{\alpha}-L_{\beta}}^\prime$ & $\phi_1$, $\phi_2$,  $\phi_4$&$ g'^2(v_1^2+4v_2^2+16v_4^2)$\\\hline
    \textbf{MIV} &      $U(1)_{B-3L_{\alpha}}$ & $\phi_3$, $\phi_6$&$ g'^2(9v_3^2+36v_6^2)$\\
    \hline
   \end{tabular}
    \caption{\footnotesize Singlet scalar fields $\phi_i$ having charges $i$ under $U(1)^\prime$.}
    \label{tab:U1pModels}
\end{table}

\section{Relic density and direct detection}\label{sec:RDandDD}
\begin{figure}[t!]
\centering
\includegraphics[height=4cm, width = 6cm ]{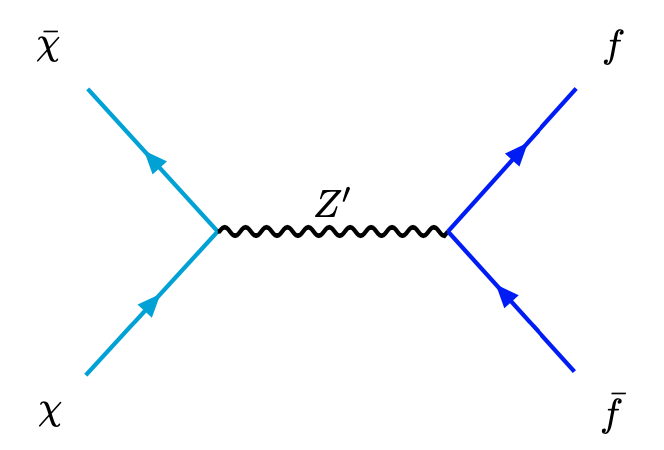}~ ~ ~ ~ ~ ~ 
\includegraphics[height=5cm, width = 4cm ]{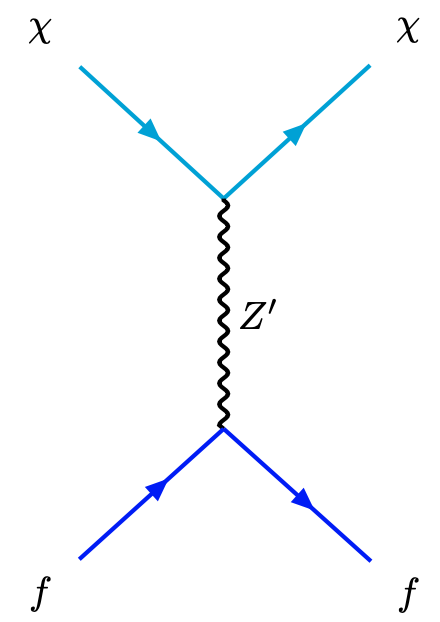}
\caption{\footnotesize Feynman diagrams leading to dark matter annihilation in the thermal \textit{freeze-out} process (left) and dark matter direct detection (right).}
\label{fig:feynmanannihilation}
\end{figure}

In this work we investigate the possibility of $\chi$ being a thermally produced WIMP dark matter candidate, taking into account experimental $Z'$ constraints to its possible couplings to the SM fields. The relic density of $\chi$ is determined by the thermal \textit{freeze-out} with resonant enhancement of the $Z'$ mediated annihilation cross-section. The kinematically allowed processes are $\overline{\chi} \chi \rightarrow \overline{f} f$ for SM fermions $f$ with masses $M_f>M_\chi$, as shown in Fig. \ref{fig:feynmanannihilation}  (see left panel). The t-channel $\overline{\chi} \chi \rightarrow Z' Z'$ process is kinematically forbidden by the resonance condition $2 M_\chi \sim M_{Z'} $. In the parameter space considered here, the leading contributions to the relic density determination are the annihilation into neutrinos, followed by charged leptons, while quarks contribute $\mathcal{O}(1-10 \%)$. The resonant condition allows lower values of $g'$ compared to the non-resonant case down to $g'\sim 10^{-6}$ for a $50$ MeV $Z'$ mass, for example. We filter the data to reproduce the observed relic density $\Omega_{\rm CDM}h^2=0.1198$~\cite{Planck:2018vyg}. 

The gauge coupling of DM to the $Z'$ leads to tree level spin-independent (SI) scattering of DM with nucleons. The Feynman diagram corresponding to this process is shown in Fig. \ref{fig:feynmanannihilation}  (see right panel). The most stringent limits on the elastic SI nucleon-DM scattering to date have been obtained by the Xenon1T \cite{XENON:2020gfr} and PandaX-II \cite{PandaX-II:2017hlx} collaborations. The SI scattering limits weaken as the DM mass drops below $\sim 20$ GeV. Future experiments are projected to explore this low region mass, down to the neutrino floor. One such experiment is the SBC-DM \cite{Giampa:2021wte} in its future dark matter 1 ton-yr phase. We consider both current Xenon1T and PandaX-II results as well as the 1 ton-yr SBC-DM projection. 

The dark matter-nucleus SI cross-section per nucleon, in the small momentum transfer limit, is given by \cite{Alves:2015pea}
\begin{align}\label{eq:SI}
\sigma_{\rm SI} \approx \frac{\mu^2_{\chi n}}{\pi}  \frac{( Z f_{p} + (A-Z) f_{n})^2}{A^2} \;,
\end{align}
where, $\mu_{\chi n}$  is the WIMP-nucleon reduced mass, $Z$, $A$ are the atomic number, atomic mass of the target nucleus, respectively.
Also,  $f_{p}$ and $f_{n}$ represent proton and neutron  scattering functions and are given by
\begin{equation}
f_{p}\approx  \dfrac{g_{\chi}}{M^{2}_{Z^{\prime}} }  (2g_u + g_d),   ~ ~ ~  f_{n }\approx  \dfrac{g_{\chi}}{M^{2}_{Z^{\prime}} }  (g_u + 2g_d) \;.
\end{equation}
In this study, $g_u = g_d = g_\chi= g^\prime/3$, hence $f_p = f_n \approx  \dfrac{g^{\prime 2}}{3 M^{2}_{Z^{\prime}} } $.  Therefore,  the spin-independent cross-section  reduces to
\begin{equation}\label{eq:SI1}
\sigma_{\rm SI} \approx  \frac{\mu^2_{\chi n}}{\pi}  \dfrac{g^{\prime 4}}{{ 9 M^{4}_{Z^{\prime}} }} \;.
\end{equation}
Due to the vector-like nature of $\chi$, axial couplings to $Z'$ of the form $g_A \overline{\chi}\gamma^\mu \gamma_5 Z^\prime_\mu \chi $ are not present at tree-level. Therefore, the spin dependent scattering cross-section arises at one-loop level and is consequently subdominant. Hence, we consider only constraints coming from SI scattering experiments.
Given the resonance condition $2M_\chi\sim M_{Z^\prime}$, we can project the constraints from direct detection experiments to the $M_{Z^\prime} - g^{\prime}$ parameter space utilizing Eq. (\ref{eq:SI1}). Conversely, we can project the constraints on $M_{Z^\prime} - g^{\prime}$ from \cevns, collider, beam dump, and cosmology to the $M_{\chi} - \sigma_{SI}$ parameter space. 

\section{Dark matter direct detection and CE$\nu$NS complementarity}\label{sec:DM-CENES}
Here, we present our numerical results in search of  $Z^\prime$ boson as well as DM considering various constraints arising from both the dark matter direct detection experiments as well as experiments searching for  CE$\nu$NS processes.   Moreover,  in order to have a comprehensive understanding  of the allowed parameter space,  constraints arising from various other experiments are also presented. 
 In the left panel, exclusion regions   are shown using different color in the ($M_{Z}^\prime -  g^{\prime}$)  plane, whereas in the right-panel we show exclusion limits for ($M_{\chi} -  \sigma_{\rm SI}$)  plane.
\begin{figure}[t!]
\centering
\includegraphics[width = 0.48\textwidth ]{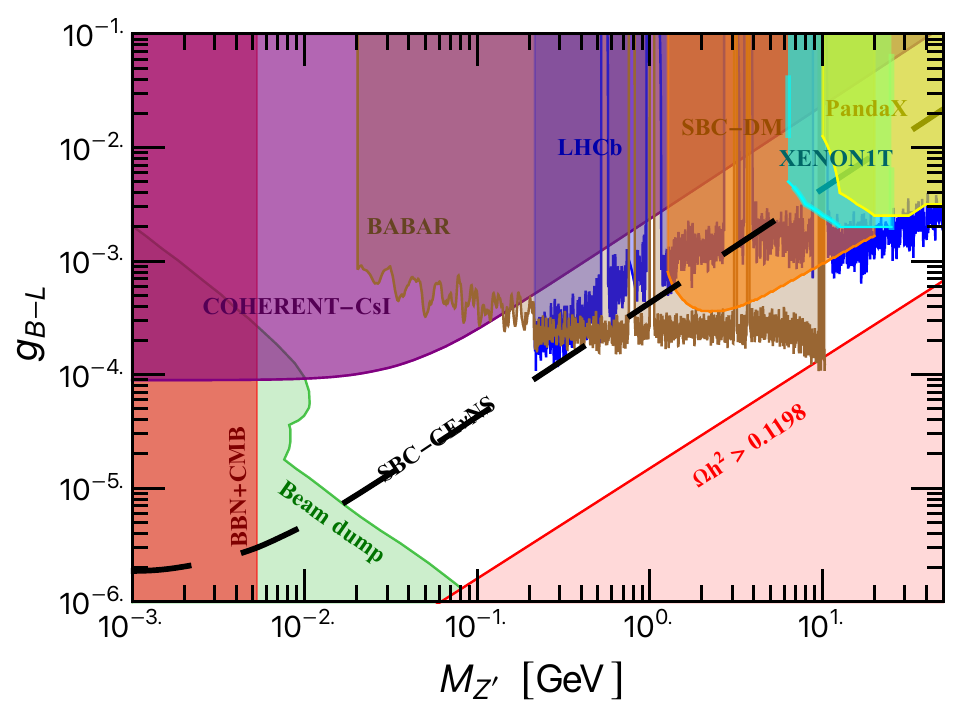}
\includegraphics[ width = 0.48\textwidth ]{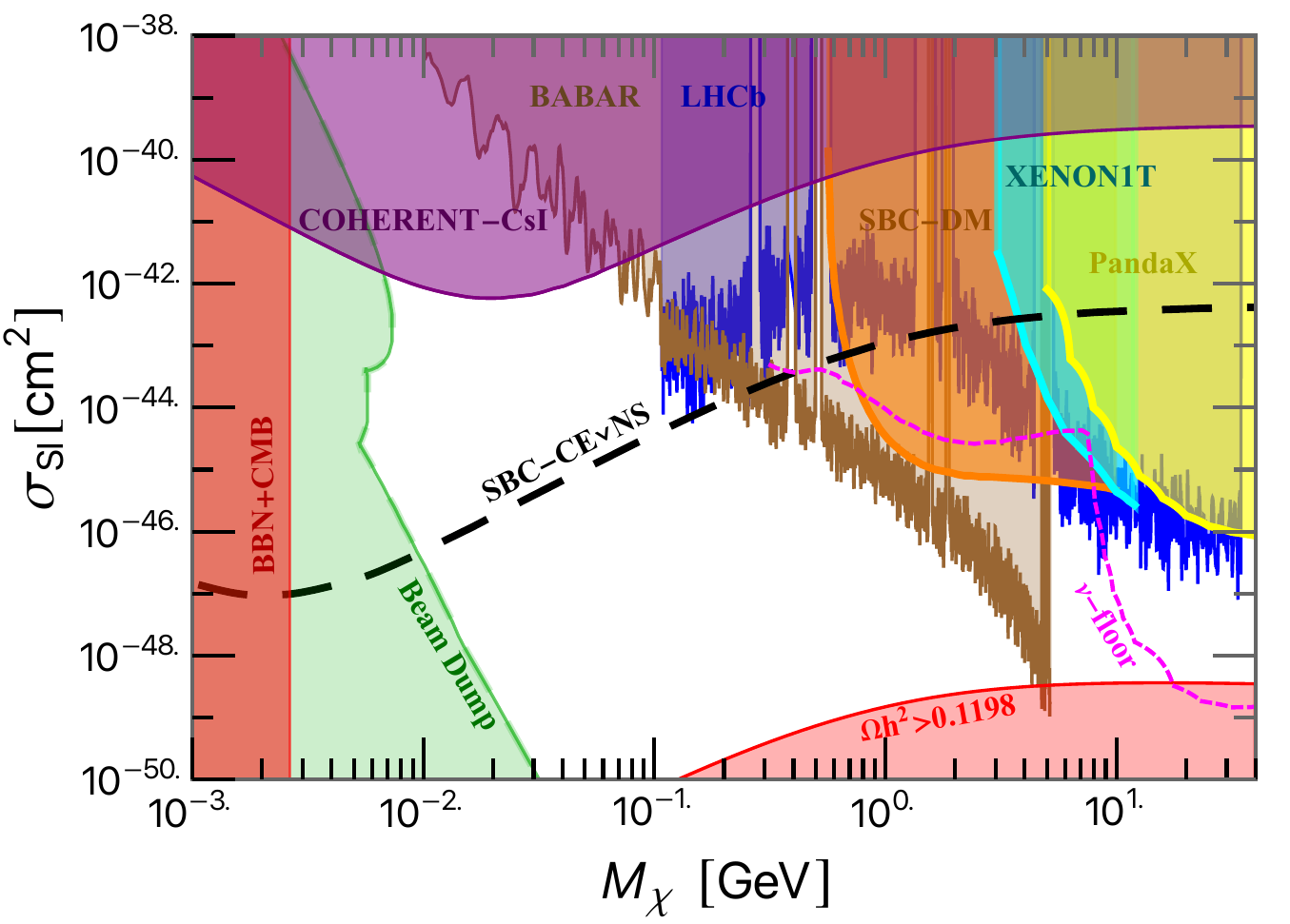}
\caption{\footnotesize Left panel: Exclusion regions in the $(M_{Z^\prime} , g^\prime)$  plane for the B-L model. Right panel: Exclusion regions for the spin independent cross-section in the  $(M_{\chi} , \sigma_{\rm SI})$ plane for the B-L model. The light-purple shaded area corresponds to the constraint set by the current COHERENT-CsI data~\cite{Akimov:2017ade}.  The limits set by the future reactor-based  \cevns experiment SBC~\cite{Giampa:2021wte,SBC:2021yal}, is presented using the black long dashed line (SBC-\cevns). 
The exclusion regions set by the BBN and CMB~\cite{Kamada:2015era},  beam dump experiments~\cite{Bergsma:1985qz,Bjorken:1988as,Riordan:1987aw,Bross:1989mp,Konaka:1986cb,Blumlein:1990ay,Banerjee:2019hmi,Astier:2001ck,Davier:1989wz,Bernardi:1985ny,Blumlein:2013cua},  BABAR \cite{Lees:2014xha} and LHCb dark photon searches~\cite{Aaij:2019bvg} are presented using the red,  light-green,  light-brown and sky-blue regions, respectively.  Limits set by the dark matter experiments are presented using  the light-orange, light-cyan, and light-yellow regions for the SBC-DM \cite{Giampa:2021wte},  XENON1T \cite{XENON:2020gfr},  and PandaX-II \cite{PandaX-II:2017hlx} experiments, respectively  (see text for more details). In the right panel, the argon $\nu$-floor background~\cite{Ruppin:2014bra} has been marked using dotted-magenta curve.
}
\label{fig:B-L}
\end{figure}
%
We start our discussion by analyzing  the flavor-independent \u1p  (i.e., $ U(1)_{B-L} $) model, which is shown in Fig. (\ref{fig:B-L}). The constraints arsing from the COHERENT-CsI data has been presented using the light-purple shaded region.  
The exclusion regions arising from the future reactor-based  \cevns  experiment SBC-\cevns~\cite{Giampa:2021wte,SBC:2021yal}  are shown using the black long-dashed  line. 
In order to present results for these  \cevns  experiments in the $(M_{Z^\prime} , g^\prime)$ plane,   we  perform a $ \chi^{2} $ test using the latest COHERENT-CsI data~\cite{Akimov:2017ade} as well as for the SBC-\cevns~\cite{Giampa:2021wte,SBC:2021yal},   following numerical procedures discussed in Sec.  \ref{sec:Cevens}, and final exclusion limits are presented at 95\% confidence level.   
Regarding the limits from dark matter direct detection experiments, we show these bounds using the light-orange, light-cyan, and light-yellow region for the SBC-DM,  XENON1T and PandaX-II experiments, respectively.
Moreover,  the light-red regions represent the exclusion limit of  the relic density  calculation  for the given model, due to an overabundance of dark matter. 

Bounds arising from the calculation of $ \Delta N_{\rm eff} $ by  the BBN + CMB~\cite{Kamada:2015era} measurements  are shown using the vertical red band.
It is to be noted that   different electron beam dump experiments like E141 \cite{Riordan:1987aw}, E137 \cite{Bjorken:1988as}, E774 \cite{Bross:1989mp},  KEK \cite{Konaka:1986cb},  Orsay \cite{Davier:1989wz}, and NA64 \cite{Banerjee:2019hmi}, which put bounds on dark photon searches, can also put bounds on masses and couplings of  \zp boson.  
 Moreover,   proton beam dump experiments like $ \nu $-CAL I \cite{Blumlein:1990ay},  proton bremsstrahlung \cite{Blumlein:2013cua}, CHARM \cite{Bergsma:1985qz}, NOMAD \cite{Astier:2001ck},  and  PS191 \cite{Bernardi:1985ny} can also set bounds on \zp boson searches. We consider these bounds from the literature.  In order to recast these limits,   \texttt{Darkcast}~\cite{Ilten:2018crw} code has been utilized to our specific model, and the combined electron and proton beam dump limits are  presented using light-green region at 90\% confidence level.

We consider limits set by  the proton-proton collider  LHCb~\cite{Aaij:2019bvg}, where \zp arising from \u1p symmetry decays to $ \mu^{+} \mu^{-} $.    Similarly,  we also consider \zp production in the  electron-positron collider BABAR \cite{Lees:2014xha}.  For both the LHCb and BABAR experiments, we have utilized  \texttt{Darkcast}~\cite{Ilten:2018crw} code  to recast their results  as shown using the sky-blue and light-brown regions at 90\% confidence level,  respectively.   Notice that for the BABAR, one  could also have bound on $ \mu^{+} \mu^{-} $ production \cite{BaBar:2016sci},   i.e.,  for scenarios where we have $ x_e = 0,   x_{\mu} \neq 0 $. However, it has been observed that those bounds are much  weaker compared to our COHERENT-CsI analysis or LHCb bounds~\cite{Aaij:2019bvg}.  Hence,  throughout this work we  only entertain bounds arising from the BABAR for  $  x_{e} \neq 0  $.

\begin{figure}[t!]
    \centering
\includegraphics[height=22cm, width = 8cm ]{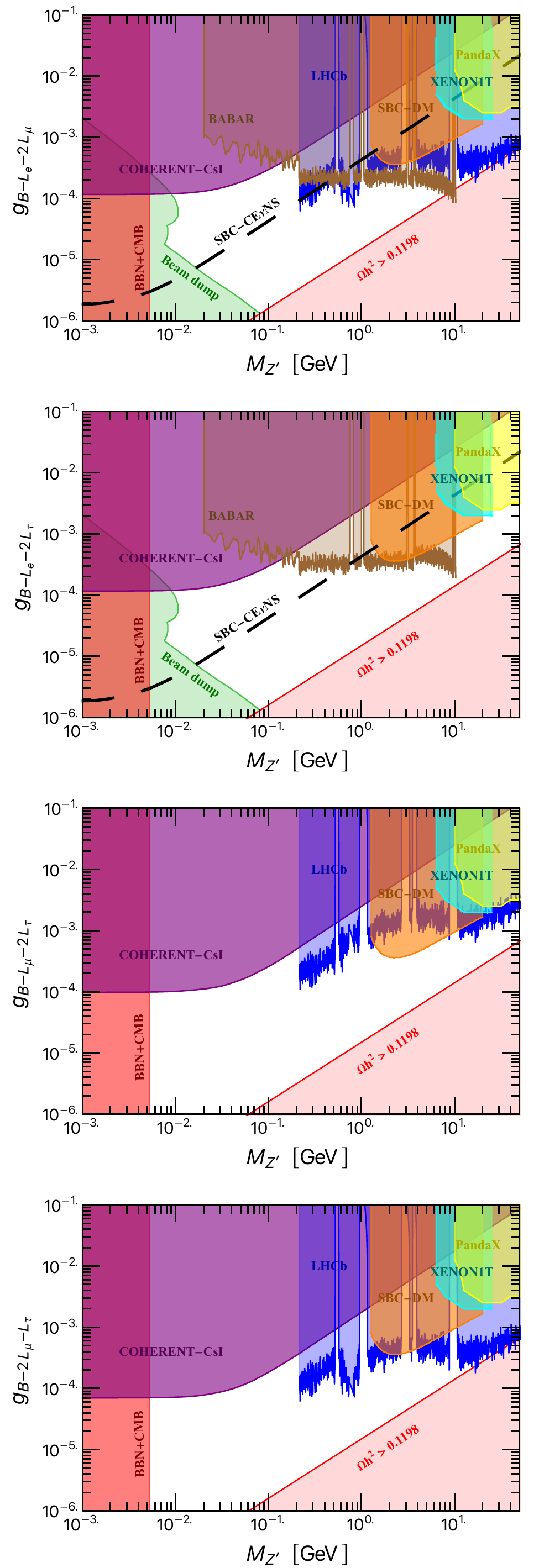}
\includegraphics[height=22cm, width = 8cm ]{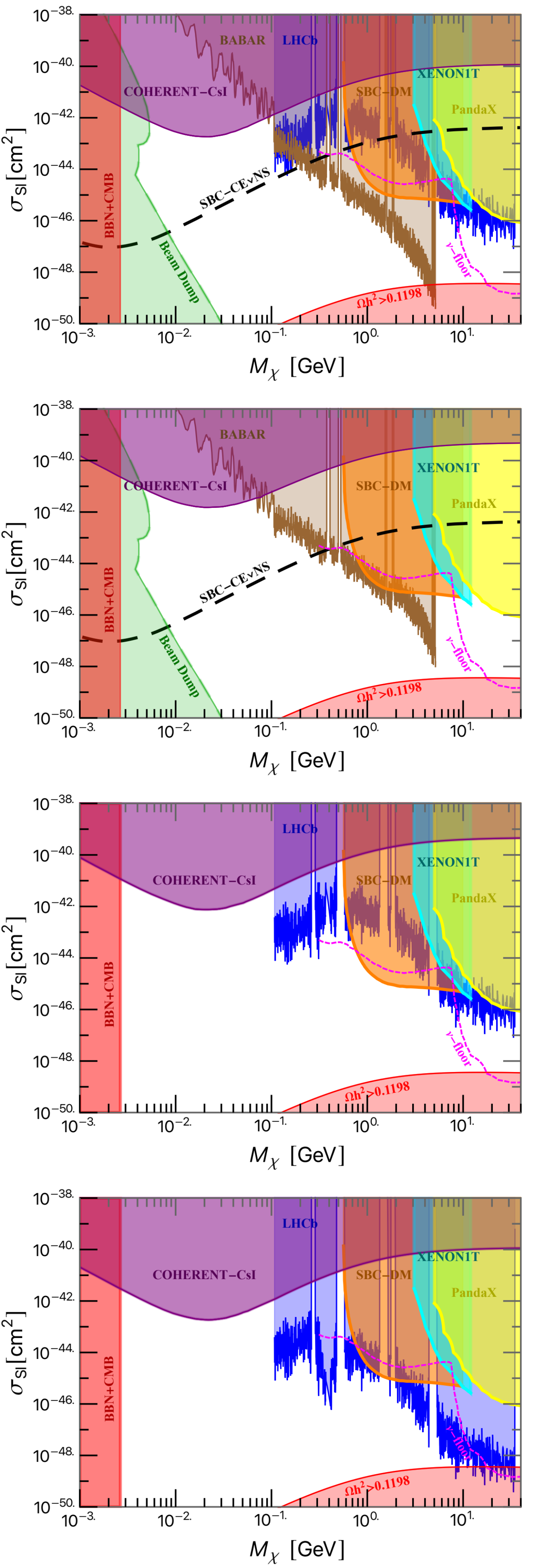}
\caption{\footnotesize Same as Fig. (\ref{fig:B-L}) but for \textbf{MII} \u1p models as given by Table (\ref{tab:U1pModels}).  }
\label{fig:FlavoredU1a}
\end{figure}

We  notice  from the left panel of Fig. (\ref{fig:B-L}) that the forthcoming \cevns experiment SBC-\cevns will be able to produce the most stringent constraint for   $Z^\prime$ masses between $(0.02-1.3)$ GeV and couplings in the range ($ 10^{-5} - 4 \times 10^{-4} $). The reason for this is the high antineutrino flux ($10^5$ times higher than the one from the SNS) given the considered 1~MW$_{th}$ power reactor and 3~m baseline~\cite{Huber:2011wv, Mueller:2011nm},  the very low energy threshold achieved by the detector, and the fact that the background is greatly reduced due to the detector insensitivity to electron recoils, for more details  see~\cite{SBC:2021yal}. It can be observed that the SBC-\cevns will be able to constrain $(M_{Z^\prime} , g^\prime)$ parameter space almost an  $\mathcal{O}(1)$ stronger than the latest constraint provided by the COHERENT-CsI data. 
For the mass range between $(0.2-1.3)$ GeV the future SBC-\cevns bounds will be competitive with the current  LHCb exclusion limits.  However,  constraint coming from the BABAR collaboration  puts the most stringent bounds in the range   $(0.5-10)$ GeV of \mzp, whereas above 10 GeV the LHCb again shows the best exclusion limits. 
Interestingly,  for coupling of \zp boson $ \sim 10^{-4} $ or smaller are concerned,  it can be seen that the BBN+CMB calculations as well as beam dump bounds  are able to ruled out a significant amount of the parameter space for \mzp less than 0.1 GeV (see shaded green and red regions). 
On the other hand, for \mzp greater  than 0.1 GeV, it is the relic density calculations that can rule out most of the parameter space for smaller \gp as shown by the light-red regions.

We observe further that  the forthcoming SBC-DM dark matter constraint will surpass the present bounds provided by XENON1T and PandaX-II, respectively. 
It is to be noted further that in this scenario, the right-handed neutrino mass matrix is generated dynamically, once the the scalar field $\phi_2$ takes it's $vev$ by breaking of $U(1)_{B-L}$ symmetry.

For the same model we also show the corresponding plot in the ($M_\chi - \sigma_{SI}$) plane in the right panel of Fig. (\ref{fig:B-L}), showing the same experimental constraints. The \cevns process from solar, atmospheric and diffuse supernova neutrinos constitute an unavoidable background for dark matter direct detection searches. We have indicated this background \cite{Ruppin:2014bra} with a dotted magenta curve in the results, as a discovery limit, not an exclusion region. We notice that the bounds from BABAR and LHCb constrain the parameter space down to the neutrino floor of direct detection experiments, for dark matter masses below $M_\chi\sim 8$ GeV.

\begin{figure}[t!]
    \centering
\includegraphics[height=11cm, width = 8cm ]{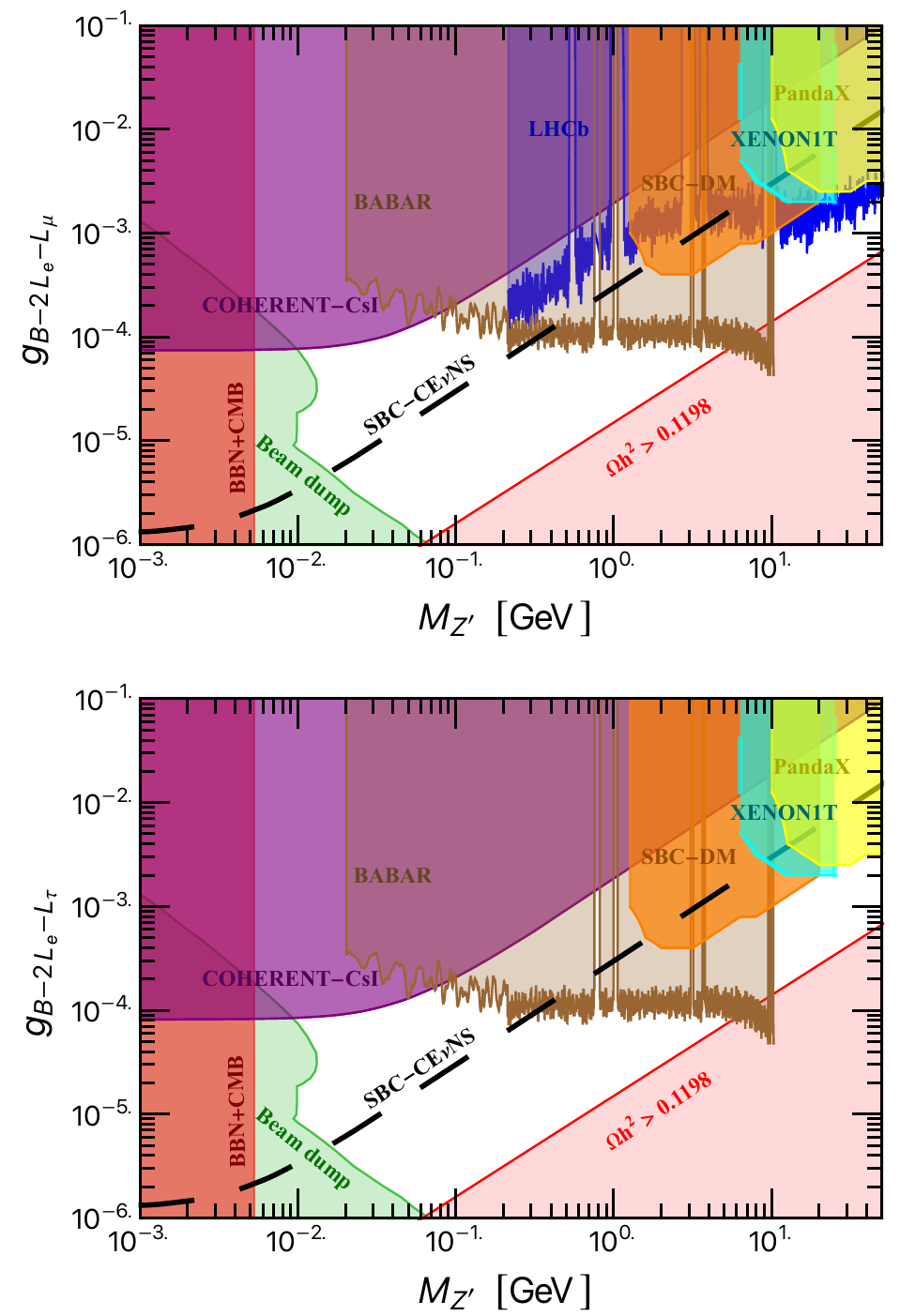}
\includegraphics[height=11cm, width = 8cm ]{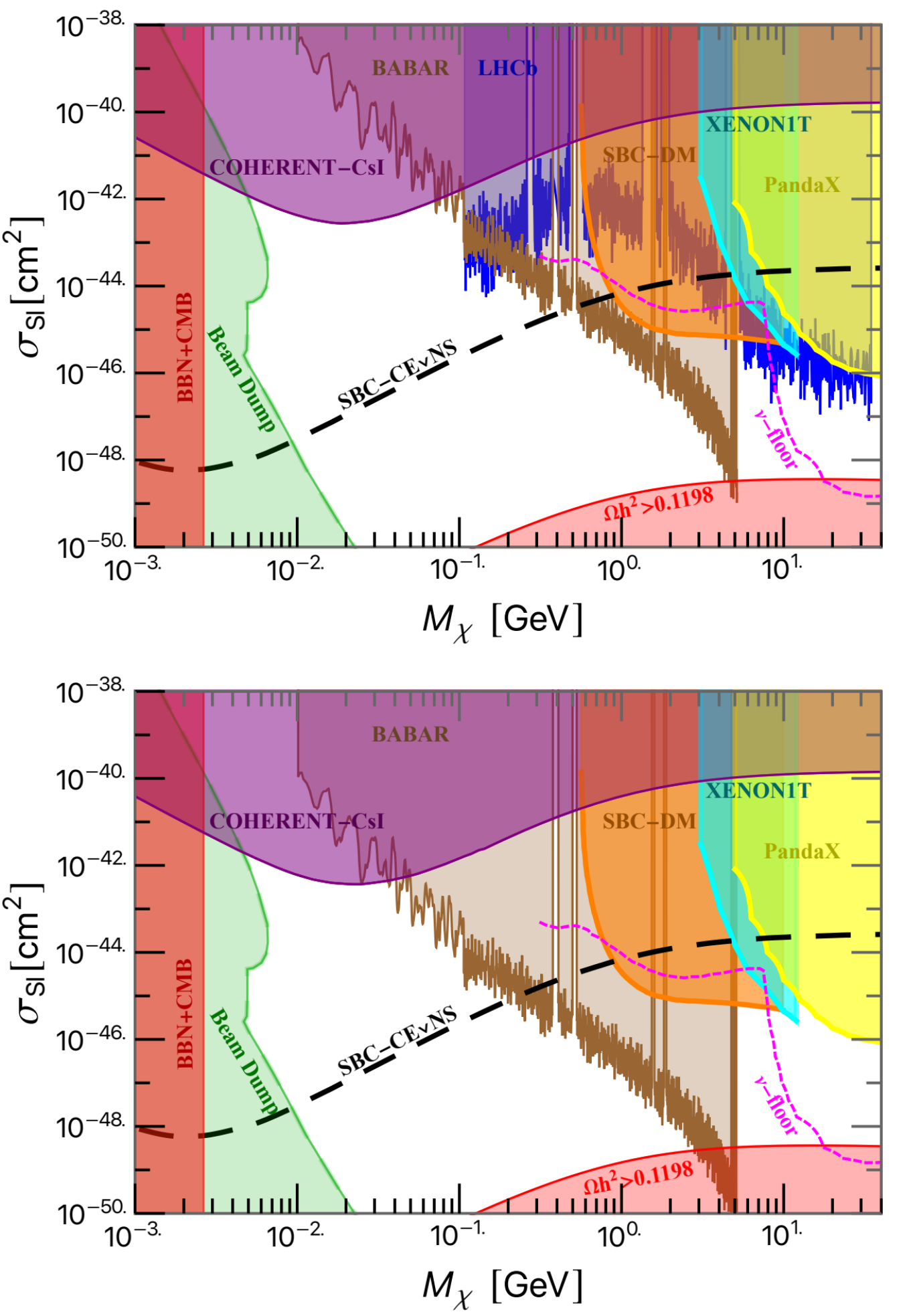}
\caption{\footnotesize Same as Fig. (\ref{fig:B-L}) but for the  \textbf{MIII}  \u1p models as given by Table (\ref{tab:U1pModels}). }
\label{fig:FlavoredU1b}
\end{figure}

Having discussed the flavor-independent B-L model, we now proceed to explore parameter space for the flavor-dependent B-L models, which are presented in Fig. (\ref{fig:FlavoredU1a}).  The four scenarios investigated here correspond to the model \textbf{MII} as given by Table (\ref{tab:U1pModels}) and depending on $U(1)^\prime$ charges (see Table (\ref{tab:modelcharges})), we have  $ U(1)_{B-L_e-2L_\mu},   U(1)_{B-L_e-2L_\tau},  U(1)_{B-L_\mu-2L_\tau}$, and  $U(1)_{B-2L_\mu-L_\tau}$,  respectively. 
It is important to point out here that these four scenarios have very sharp predictions for the leptonic sector. All these cases  lead to two-zero textures in the neutrino mass matrix,   and are consistent with the latest global-fit of neutrino oscillation data, detailed phenomenology of these cases for the leptonic sector have been carried-out in  \cite{Flores:2020lji}. 

Notice that  beam dump experiments, BABAR as well as reactor experiment SBC-\cevns  only provide bounds for the electron channel, therefore we find their impact only for the cases where one has non-zero $ x_e $ as can be seen from the first and second row  of the Fig. (\ref{fig:FlavoredU1a}). 
Similarly,  the LHCb dark-photon searches are relevant for non-zero $ x_\mu$, and hence one sees its presence for three cases except for the second row (see sky-blue areas).
In what follows, investigating  all these four panels it can be noticed that the first panel of the upper row shows the most stringently constrained region compared to the remaining cases.  For all the possible charges, the constraints from direct DM searches translate into bounds for the $M_{Z^\prime}-g^\prime$ and we can also translate the different bounds from neutrino and accelerator experiments to the DM direct searches plane.
For  cases where we have non-zero $ x_e $, there are constraints from beam dump, BABAR, COHERENT, and reactor neutrinos such as SBC-\cevns. While for non-zero  $ x_\mu$ there are bounds from COHERENT and  LHCb experiments.  Therefore, the most restricted allowed parameter space has been observed for these cases (i.e. for $x_e \neq 0 \neq x_\mu$). 
From the third and fourth row, we find that as there are no constraints from  beam dump experiments  or SBC-\cevns, most of the parameter space will remain unexplored. The unexplored regions for \zp boson masses lie in  ($ 5.3-60 $) MeV with the coupling constant below $ 10^{-4} $ as shown in the left panel. For DM masses, most of the unconstrained regions lie in ($2.8-130$) MeV (see right panel). 

\begin{figure}[t!]
    \centering
\includegraphics[height=18cm, width = 8cm ]{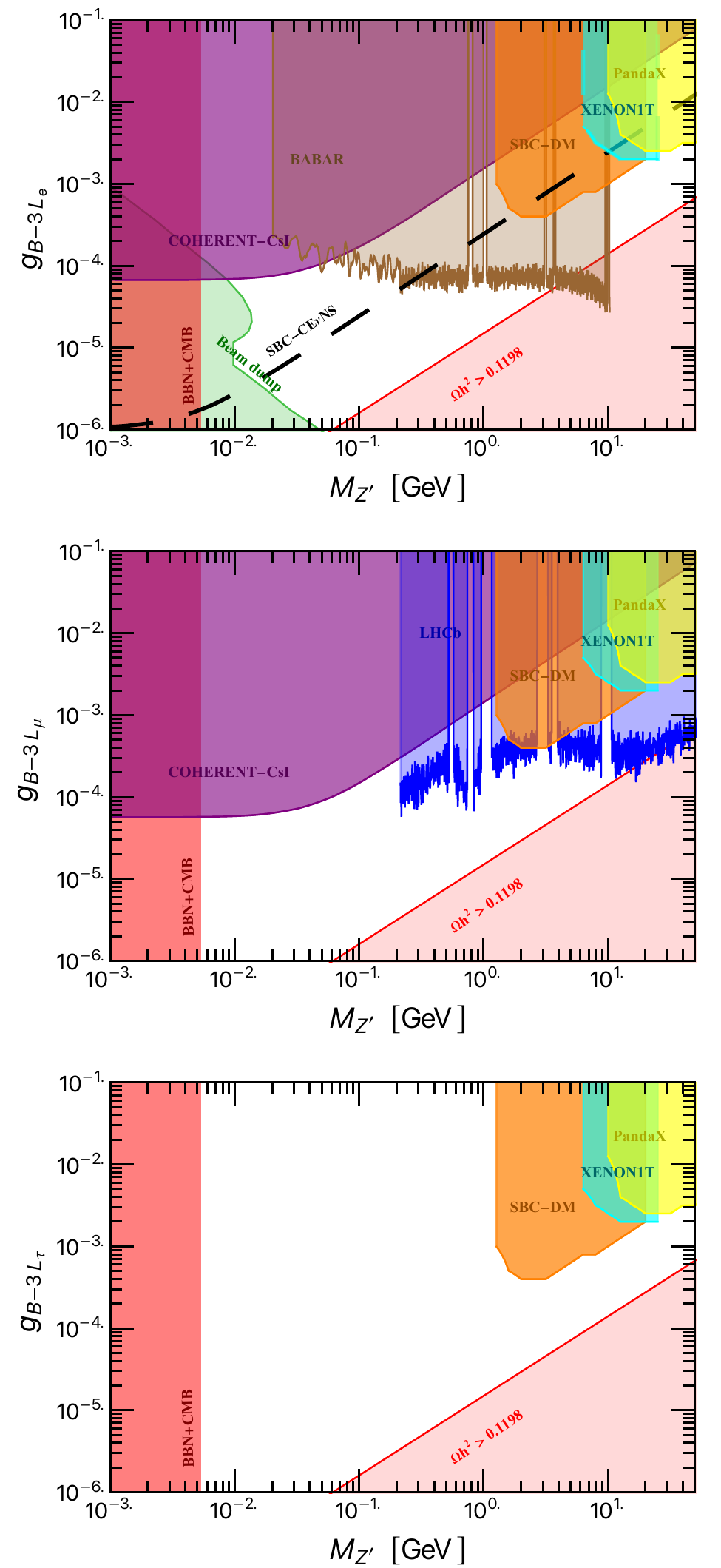}
\includegraphics[height=18cm, width = 8cm ]{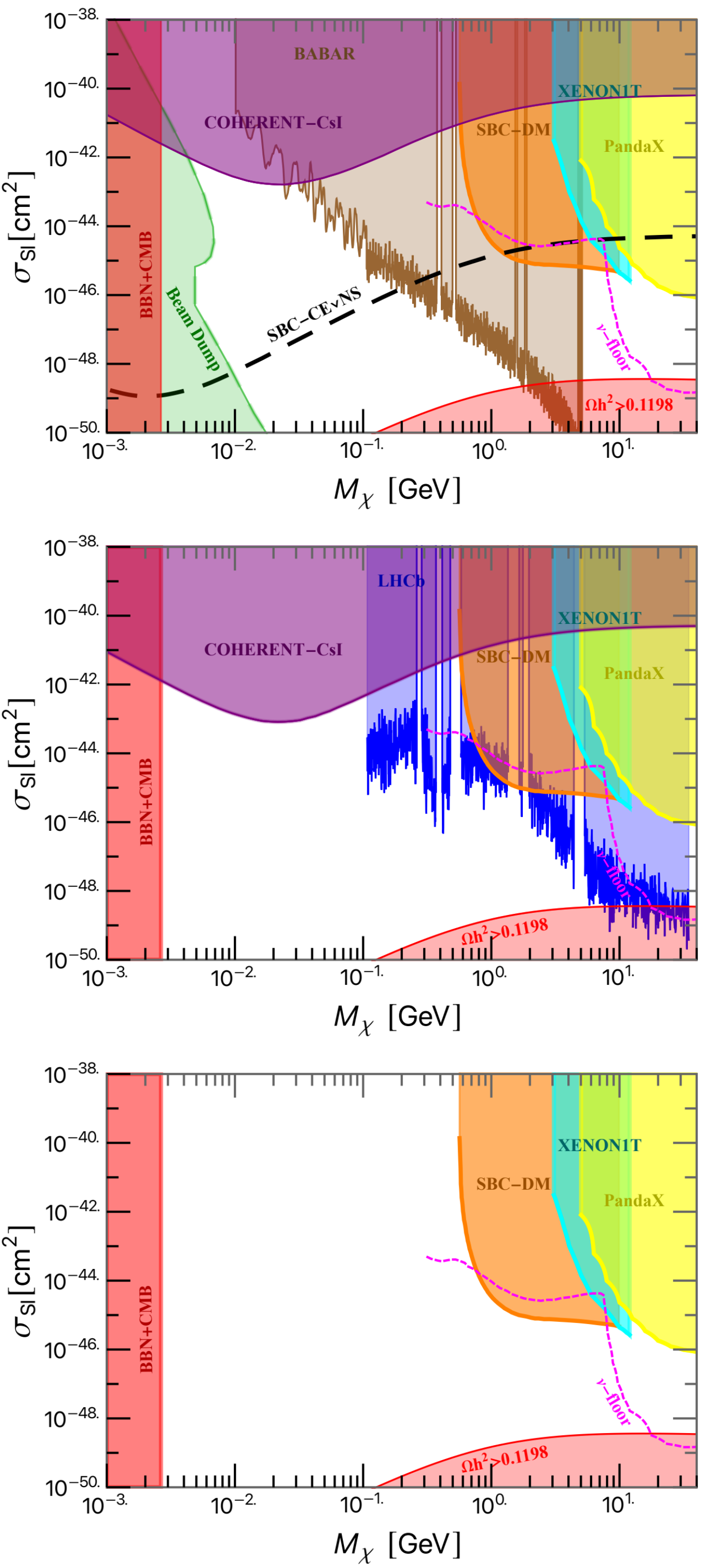}
\caption{\footnotesize Same as Fig. (\ref{fig:B-L}) but for \textbf{MIV} \u1p models as given by Table (\ref{tab:U1pModels}). }
\label{fig:FlavoredU1c}
\end{figure}

In Fig. (\ref{fig:FlavoredU1b}), we discuss our numerical results for \u1p models corresponding to the model  \textbf{MIII}  as given by Table  (\ref{tab:U1pModels}).   
All scenarios of  models \textbf{MIII} have  correlations for the neutrino masses and mixing parameters, see appendix \ref{app:MIV}, and are compatible with the latest neutrino data.

 As per the  constrained parameter space is concerned, Fig. (\ref{fig:FlavoredU1b}) shows almost similar pattern like Fig. (\ref{fig:FlavoredU1a}).  Here, we notice that the first row shows mildly better  constraint compared to second row because of the presence of LHCb for the mass range above 10 GeV. 

 Our simulated results for models $ U(1)_{B-3L_e},   U(1)_{B-3L_\mu},  $  and  $ U(1)_{B-3L_\tau}$,  are presented in Fig.  (\ref{fig:FlavoredU1c}),  in the first, second and third rows respectively.   
 Also, all cases of models \textbf{MIV} contains enough freedom to reproduce neutrino masses and mixing which are in good agreement with the latest neutrino data,  but predict no correlations.
 Here again one observes that the first row shows the most restricted parameter space.

 From the left panel of the second row it can be seen that the  most of the parameter space in the range ($ 5.3-60 $) MeV for \gp below $ 10^{-4} $ remains unexplored. 
 On the other hand,  the left panel of the third row does not show any constraint in   ($ 5.3\times 10^{-3}-0.1$) GeV mass range of \zp,   whereas  above 0.1 GeV  for coupling constant below $ 10^{-3} $,  the relic density calculation explored all the parameter space. 
 Also,   above 10 GeV of \mzp mass with \gp greater than $ 3\times 10^{-4} $,   it is the forthcoming SBC dark matter searches that can able to put the most strong bounds on the allowed parameter space.  A similar conclusion holds true for the right panel, but for the DM mass in  ($2.8-130$) MeV. 
\section{Conclusions}\label{sec:conclusion}
We have discussed scenarios for a light gauge boson $Z^\prime$, where the SM fermions and the DM interact with such a gauge boson. In these scenarios, the DM stability  is due to a residual symmetry from the spontaneous breaking of $U(1)^\prime\rightarrow Z_N$. 
We have found that in order to have the correct thermal DM relic density, the DM annihilation in the early Universe must occur resonantly,  i.e.  DM mass should satisfy $M_{\chi}\approx M_Z^\prime/2$.  Firstly,  the flavor-independent \u1p model i.e.,  $B-L$ scenario has been studied, where all  leptons have the same charge. Later,  different flavorful scenarios have also been analyzed, where different lepton flavors carry different \u1p charges. 
Since the quarks also transform under  the extra gauge symmetry, the $Z^\prime$ couples to the nucleons, and therefore,  we have investigated  constraints arising from  CE$\nu$NS (when the  electron and/or muon flavors are charged) as well as from DM direct detection experiments.   
However,  it goes without saying that any new physics scenario beyond the SM undergoes various phenomenological constraints coming from numerous particle physics experiments.    Using various other limits, coming from the beam dump, LHCb,  and BABAR, we have first constrained   $M_{Z^\prime}-g^\prime$ allowed parameter regions.  
Later, we have translated all these limits in the  $M_{\chi}-\sigma_{\rm SI}$ plane using the correlation between the DM mass ($ M_{\chi} $) ,  spin-independent DM direct detection cross-section ($ \sigma_{\rm SI} )$ together with gauge coupling (\gp).

Our noteworthy results are summarized in Figs. (\ref{fig:B-L},  \ref{fig:FlavoredU1a},  \ref{fig:FlavoredU1b},  and \ref{fig:FlavoredU1c}).
Investigating all scenarios, it has been observed that the most stringent parameter space is obtained for the \u1p model with $ x_e \neq 0 $.    
For the particular scenario, we have bounds from the \cevns  , beam dump, and BABAR together with DM direct detection experiments.  
As an example, from the first panel of Fig. (\ref{fig:FlavoredU1c}) it can be noticed that the forthcoming reactor-based \cevns experiment SBC-\cevns can explore parameter space with gauge coupling of $ \mathcal{O} (10^{-5}) $ or smaller for \zp masses around 0.1 GeV. 
Also, we have observed that the future SBC-\cevns can put almost an order of magnitude stronger constraint compared to the latest COHERENT-CsI bound. 
On the other hand,  the most unexplored  regions are seen for scenarios with $ x_{e} = 0,  x_{\mu} = 0$, as there are no bounds from the \cevns, beam dump,  LHCb,  or BABAR.  In this case,   in the \zp mass range ($ 5.3\times 10^{-3}-0.1$) GeV parameter space remains unconstrained. 
From the DM relic density,  there is  a lower bound in the $M_{Z^\prime}-g^\prime$ plane due to kinematics. All these constraints can be translated to the DM direct detection cross-section setting strong constraints for DM masses below 10 GeV for PandaX-II,  6 GeV for XENON1T, whereas the stringent constraint is observed for the SBC-DM searches, which puts a limit around  1.3 GeV. 
Finally,  it is worth mentioning that for DM mass below 1.3 GeV there are no constraints from the current experiments, and in this way, there is a complementarity in bounds from CE$\nu$NS and DM direct searches.

\acknowledgements
This work is supported by 
the  German-Mexican  research  collaboration grant SP 778/4-1 (DFG) and 278017 (CONACYT),  
the grants CONACYT CB-2017-2018/A1-S-13051 (M\'exico), the DGAPA UNAM grant PAPIIT IN107621 and SNI (M\'exico). NN is supported by the postdoctoral fellowship program DGAPA-UNAM. LMGDLV is supported by CONACYT. The work of LJF is supported by a CONACYT postdoctoral fellowship.

\appendix
\section{Correlations in the \textbf{MIII} models}
 \label{app:MIV}
 In the models $B-2L_e-L_{\mu}$ and $B-2L_e-L_{\tau}$, the charged lepton and Dirac neutrino mass matrices are diagonal by construction. While it is not enough to include the $\phi_1$ and $\phi_2$ to reproduce neutrino masses and mixings~\cite{Flores:2020lji}. This can be alleviated by the inclusion of a third flavon field $\phi_4$. In this way the right-handed neutrino mass matrix takes the form
 \begin{equation}
     M_{B-2L_e-L_\mu}=\left(\begin{array}{ccc}
     y_4 v_4&0&y_2 v_2\\0&y_3 v_2&y_1 v_1\\y_2 v_2& y_1v_1& M_1
     \end{array}\right), 
     M_{B-2L_e-L_\tau}=\left(\begin{array}{ccc}
     y_4 v_4&y_2 v_2&0\\y_2 v_2&M_1&y_1 v_1\\0& y_1v_1& y_3 v_2
     \end{array}\right) \;,
     \label{eq:MIIIRHN}
 \end{equation}
 for  $U^{\prime}_{B-2L_e-L_{\mu}}$, $U^{\prime}_{B-2L_e-L_{\tau}}$  models, respectively. 
 Given the Dirac and Majorana neutrino mass matrices, one can construct the low-energy active neutrino mass matrices using type-I seesaw mechanism. We write the active neutrino mass matrix as
 \begin{equation}
     m_{\nu}=\left(\begin{array}{ccc}
     m_{11}&m_{12}&m_{13}\\ 
     * &m_{22}&m_{23}\\
      * & * &m_{33}
     \end{array}\right)\;.
 \end{equation}
From the parameters in Eq. (\ref{eq:MIIIRHN}), we find following correlations 
 \begin{align}
     m_{12}m_{33} & =m_{13}m_{23} \;, \nonumber \\
     m_{12}m_{23} & =m_{13}m_{22}\;,
 \end{align}
for $U(1)_{B-2L_e-L_\mu}$ and $U(1)_{B-2L_e-L_\tau}$, respectively. We have found that these correlations are compatible with current oscillation data, with the correct choice of neutrino masses and Majorana phases.
For the \textbf{MIV} models there are no correlations in the mass matrix, there is enough freedom to fit neutrino oscillation data.

\bibliographystyle{utphys}
\bibliography{references}

\end{document}